\documentclass[lettersize,onecolumn]{IEEEtran}

\usepackage{amsmath,amsfonts}
\usepackage{algorithmic}
\usepackage{algorithm}
\usepackage{array}
\usepackage[caption=false,font=normalsize,labelfont=sf,textfont=sf]{subfig}
\usepackage{textcomp}
\usepackage{stfloats}
\usepackage{url}
\usepackage{verbatim}
\usepackage{graphicx}
\usepackage{cite}
\usepackage{caption}
\usepackage{subcaption}
\usepackage{xcolor, color, soul}
\usepackage{xfrac}
\usepackage{soul} 
\usepackage[normalem]{ulem} 
\usepackage{amsthm}

\usepackage{etoolbox} 

\usepackage{subcaption}  
\usepackage{graphicx} 
\usepackage[utf8]{inputenc}
\PassOptionsToPackage{caption=false}{subfig}
\usepackage{tabularx} 
\usepackage{booktabs} 

 \usepackage[compact]{titlesec}         
 \titlespacing{\section}{2pt}{2pt}{2pt} 

\AtBeginDocument{%
  \setlength{\abovedisplayskip}{3pt}%
  \setlength{\belowdisplayskip}{3pt}%
}

\begin{document}

\title{Outage Probability Analysis of NOMA‑Enabled Hierarchical UAV Networks with Non-Linear Energy Harvesting}

\author{\IEEEauthorblockN{Faicel Khennoufa$^{1}$,
Khelil Abdellatif$^{2}$,  
Metin Ozturk$^{3}$,
Halim Yanikomeroglu$^{4}$, Safwan Alfattani$^{5}$}

\IEEEauthorblockA{$^{1}$Ecole Nationale Supérieure des Technologies Avancées, Department of Computer Science, Algiers, Algeria.}

\IEEEauthorblockA{$^{2}$LGEERE Laboratory, Department of Electrical Engineering, University of El-Oued, El-Oued, Algeria}

\IEEEauthorblockA{$^{3}$Electrical and Electronics Engineering, Ankara Yıldırım Beyazıt University, Ankara, Türkiye}

\IEEEauthorblockA{$^{4}$NTN Lab, Department of Systems and Computer Engineering, Carleton University, Ottawa, Canada}

\IEEEauthorblockA{$^{5}$King AbdulAziz University, Saudi Arabia}

}   
\maketitle
\begin{abstract}

Uncrewed aerial vehicles (UAVs) are expected to enhance connectivity, extend network coverage, and support advanced communication services in sixth-generation (6G) cellular networks, particularly in public and civil domains. Although multi-UAV systems enhance connectivity for IoT networks more than single-UAV systems, energy-efficient communication systems and the integration of energy harvesting (EH) are crucial for their widespread adoption and effectiveness. In this regard, this paper proposes a hierarchical ad hoc UAV network with non-linear EH and non-orthogonal multiple access (NOMA) to enhance both energy and cost efficiency. The proposed system consists of two UAV layers: a cluster head UAV (CHU), which acts as the source, and cluster member UAVs (CMUs), which serve as relays and are capable of harvesting energy from a terrestrial power beacon. For the considered IoT network architecture, the outage probability expressions of ground Internet of things (IoT) devices, each CMU, and the overall outage probability of the proposed system are derived over Nakagami-$m$ fading channels with practical constraints such as hardware impairments and non-linear EH. 
We compare the proposed system against a non–EH system, and our findings indicate that the proposed system outperforms the benchmark in terms of outage probability.


\end{abstract}


\section{Introduction}

\IEEEPARstart{U}{ncrewed} aerial vehicles (UAV) communications are pivotal in next-generation wireless networks due to their mobility, low cost, and capacity to cover remote areas~\cite{10938203}. Ad hoc UAV networks provide flexible wireless connectivity with extensive coverage and quick mission adaptability. However, energy scarcity poses a significant challenge, limiting mission duration and effectiveness, necessitating the development of sustainable and cost-effective solutions~\cite{11029408,10978385,karaman2025solutions}.
To mitigate this, wireless energy harvesting (EH) leverages ambient radio frequency (RF) signals to provide sustainable energy, enhancing network lifespan and reliability~\cite{10806653}. Additionally, ensuring stable communication is crucial; non-orthogonal multiple access (NOMA) enhances spectral efficiency over orthogonal multiple access (OMA) while requiring careful interference management~\cite{11029408}, as it allows multiple users to share the same spectrum simultaneously, improving connectivity. Thus, EH and NOMA enable sustainable, efficient, and resilient UAV communication networks for next-generation wireless communications.

The application of UAVs in wireless networks has attracted a lot of attention recently, inspiring various studies on improving coverage, security, capacity, and reliability \cite{10843775,ahmed2025toward}. 
The authors of \cite{liu2019placement} jointly optimized the placement and power allocation to improve the performance of the NOMA-UAV network. 
In ~\cite{pi2025outage}, a UAV-assisted dual-hop radio frequency–underwater wireless optical communication system employing NOMA was investigated, and exact as well as asymptotic expressions for the outage probability were derived under both decode-and-forward (DF) and amplify-and-forward (AF) relaying schemes. The authors of \cite{bepari2026uav} analyzed a UAV-assisted DF NOMA network in terms of outage probability over a Rayleigh fading channel.
The authors of~\cite{zhang2025joint} optimized the task offloading ratios, user associations, and service caching under constraints related to marine Internet of things (IoT) devices, UAV energy, and cache capacity.

Moreover, multi-UAV systems have gained significant attention due to their ability to coordinate missions and enhance communications efficiency in smart environments. In~\cite{10379154}, a multi-hop UAV relay network was proposed to serve ground heterogeneous users lacking direct links, with optimized resource allocation and deployment to meet diverse user needs and reduce outage probability. In \cite{lau2023general}, a UAV swarm network was analyzed regarding outage probability and capacity by characterizing the effects of UAV-to-UAV interference.
Among the various schemes proposed for multi-UAV systems the hierarchical scheme stands out as a prominent approach~\cite{10006695}. In this approach, UAVs are organized into functional layers, which typically include high-altitude leader UAVs and low-altitude follower UAVs. The leader UAVs handle coordination and communication control, while the follower UAVs perform data collection or relaying tasks~\cite{10006695}. A hierarchical UAV networks provide greater coverage and efficiency by delegating tasks between leader and follower UAVs, thereby reducing power consumption and latency, and enhancing reliability, flexibility, and scalability in large-scale, mission-critical conditions~\cite{10006695}.

On the other hand, energy conservation is a major challenge facing UAVs, especially for long-duration missions, due to limited battery capacity and high energy demands of both flight and wireless communications~\cite{ahmed2025toward}. Therefore, wireless EH has been proposed in the literature as one of the potential solutions to address the energy consumption challenge~\cite{OJHA2023101820}. 
In \cite{9707780}, the authors examined the use of RF EH based on ground sources in Internet of things (IoT) systems for UAV-assisted NOMA-based mobile-edge computing. They obtained closed-form expressions for the successful computation and energy consumption probabilities over the Nakagami-$m$ fading channel. 
The outage probability of UAV-assisted simultaneous wireless information and power transfer (SWIPT) and multiple-input multiple-output (MIMO) with NOMA systems has been studied in different contexts; in~\cite{10902111}, a multiple-input single-output (MISO)-NOMA framework for agricultural IoT was analyzed, while in~\cite{CAGIRAN2025155713}, a system with imperfect successive interference cancellation was investigated.
A NOMA-assisted full-duplex cooperative UAV-to-UAV communication system was considered in~\cite{10928333}, where the UAVs employ an EH framework. The outage expressions for UAV-to-UAV and cellular links were derived through asymptotic analysis~\cite{10928333}.

Furthermore, over the last decade, a great amount of effort has been devoted to developing integrated, cost-effective, and energy-efficient RF transceivers. However, these systems suffer from inevitable RF front-end impairments due to component mismatches and manufacturing defects. 
The authors in \cite{10510457} explored the impact of hardware impairments (HWI) and imperfect channel state information on the mean square error of over-the-air computation using UAVs. 
The authors of~\cite{10146460} studied a UAV-enabled IoT network with RF EH and DF relaying under HWI in terms of outage probability, system throughput, and energy efficiency, where the UAV collects energy for IoT devices and forwards their data to the ground base station.

{\color{black}Although previous research introduces the idea of harvesting energy from ground sources, for example, in~\cite{ahmed2025toward,OJHA2023101820,9707780,10902111,CAGIRAN2025155713,10928333}, they considered linear EH, which is an unrealistic assumption, since numerous experiments on real electromagnetic circuits have demonstrated the non-linearity of their input and output characteristics. In addition, most existing works on UAV-assisted communications consider either a single-UAV system or a flat multi-UAV system, while assuming ideal HWI and either ignoring EH or adopting linear EH systems~\cite{10843775,ahmed2025toward,liu2019placement,pi2025outage,bepari2026uav,zhang2025joint,10379154,lau2023general,10006695}. Although wireless EH techniques with or without NOMA have been analyzed for a UAV-assisted system individually~\cite{ahmed2025toward,OJHA2023101820,9707780,10902111,CAGIRAN2025155713,10928333}, a hierarchical system of multiple UAVs using NOMA and non-linear EH and HWI is not explored in the existing literature, to the best of the authors’ knowledge. 
In practical RF EH techniques, non-linear EH is a major factor, and HWIs in transceivers are inevitable components in wireless communication systems. However, analysis of non-linear EH and HWIs is rarely explored in the analysis of multi-UAV-assisted wireless communication networks. Motivated by these limitations, this work investigates a hierarchical UAV-assisted wireless communication system incorporating NOMA, non-linear EH, and HWIs, providing a more practical framework for the analysis of multi-UAV-assisted wireless communication networks.}
The system comprises two levels of UAVs: a cluster head UAV (CHU) acting as the source, and cluster member UAVs (CMUs) that harvest energy from the power beacon (PB) transmitted energy and utilize this harvested energy to relay information to ground IoT devices. We employ the NOMA technique to enhance spectrum efficiency while considering the effects of HWI on the proposed system with EH in a practical scenario. 
{\color{black}The} main contributions of this paper are given as follows:

\begin{itemize}
    \item We propose a joint design that integrates non-linear EH from PB signals with NOMA, offering enhanced spectrum efficiency and sustainable energy management in hierarchical ad hoc UAV networks. To ensure a more realistic and practical evaluation, we also consider the impact of HWI on the system.
   
    \item We derive the outage probability of the proposed hierarchical ad hoc UAV network that incorporates non-linear EH and NOMA in the presence of HWI. Furthermore, we obtain the outage probability of the system for a single CMU (called CMU outage probability), as well as the overall outage probability of a hierarchical ad hoc UAV network (i.e., network outage probability).
   
   \item To demonstrate the superiority of the proposed system, we compare it against two benchmark schemes: an ad hoc hierarchical UAV network using non-linear EH with NOMA, and a ground-based EH system where UAVs harvest energy directly from terrestrial transmitters.

\end{itemize}


The rest of the paper is organized as follows: The proposed hierarchical ad hoc UAV network with non-linear EH and NOMA under the impact of HWI is presented in Section II. We derive the outage probability of the proposed system in Section III. Finally, Section IV presents and discusses the simulation results, while Section V offers the conclusion.






\section{System Model}
As illustrated in Fig. 1, we consider a hierarchical ad hoc UAV network designed to serve a set of ground IoT devices. The network consists of a CHU, $L$ CMUs, indexed by $l=1, 2, ..., L$, $M$ ground IoT devices, indexed by $n=1, 2, ..., M$, and a PB. 
The CHU is assumed to serve as the communication source and resource management coordinator for the cluster members, while the cluster members  (i.e., CMUs) act as relay UAVs to forward data to the ground IoT devices.
In this scenario, we consider the following assumptions: 1) The CHU is a hovering platform at a fixed location, moving slowly enough to be considered stationary while providing coverage. 2) All nodes are equipped with one (transmit/receive) antenna. 3) Perfect CSI is available at all nodes, and the communication links experience Nakagami-$m$ fading. 4) The CMUs operate in the DF protocol and HD mode. 5) HWI is considered at all nodes. 
We assume that CMUs are equipped with batteries to harvest and store wireless energy, while the CHU is equipped with solar cells, enabling it to harvest energy and operate sustainably. It is assumed that the CMUs harvest energy transmitted by the PB using the TS protocol and store it in their batteries, which is then used to power subsequent transmissions to ground IoT devices.  






Considering the channel scenarios in~\cite{10006695}, the link between the CHU and CMUs, and CMUs in the UAV ad hoc network may establish a direct air-to-air (A2A) connection, which can be considered a line-of-sight (LoS) link altogether. In contrast, the link between the CMUs and ground IoT devices, PB and ground CMUs may establish an air-to-gound (A2G) connection, which can be considered a mixed LoS and non-LoS (NLoS). Thus, the path-loss coefficient of the A2A and A2G links can be formulated, respectively, as~\cite{10836898,10006695}
\begin{equation}
\Upsilon_{l}=  \zeta_{0}  d_{l}^{-\beta}
\end{equation}
and
\begin{equation}
\Upsilon_{\varpi}= \left( \zeta_{1} \Lambda_{\text{LoS}} + \zeta_{2} \Lambda_{\text{NLoS}} \right) d_{\varpi}^{-\beta}, \varpi=\{n, e\},
\end{equation}
where $\zeta_{0}$, $\zeta_{1}$, and $\zeta_{2}$ are the additional path-loss of the channel under LoS and NLoS transmission conditions, $\beta$ is the path-loss exponent of the channel. $d_{l}$ is the Euclidean distance between the CHU and CMUs.
$d_{\varpi}$ is the Euclidean distance between the CMUs and ground IoT devices, and the PB and CMUs. $\Lambda_{\text{LoS}}$ and $\Lambda_{\text{NLoS}}$ are the LoS and NLoS probabilities. Therefore, the probability of LoS channel occurrence can be written as 
\begin{equation}
\Lambda_{\text{LoS}}=\frac{1}{1+ \xi_{1} \exp{\left(-\xi_{2} \left[\arcsin{(\frac{H}{d_{n}})-\xi_{1}} \right]\right)}}, 
\label{eq:3}
\end{equation}
where $H$ is the altitude of the CMUs, $\xi_{1}$ and $\xi_{2}$ are constant values depending on the environment~\cite{10006695}. Accordingly, the probability of NLoS can be given as
\begin{equation}
\Lambda_{\text{NLoS}}=1-\Lambda_{\text{LoS}}. 
\end{equation}


\begin{figure}[]
\centering
\includegraphics[width=0.9\columnwidth]{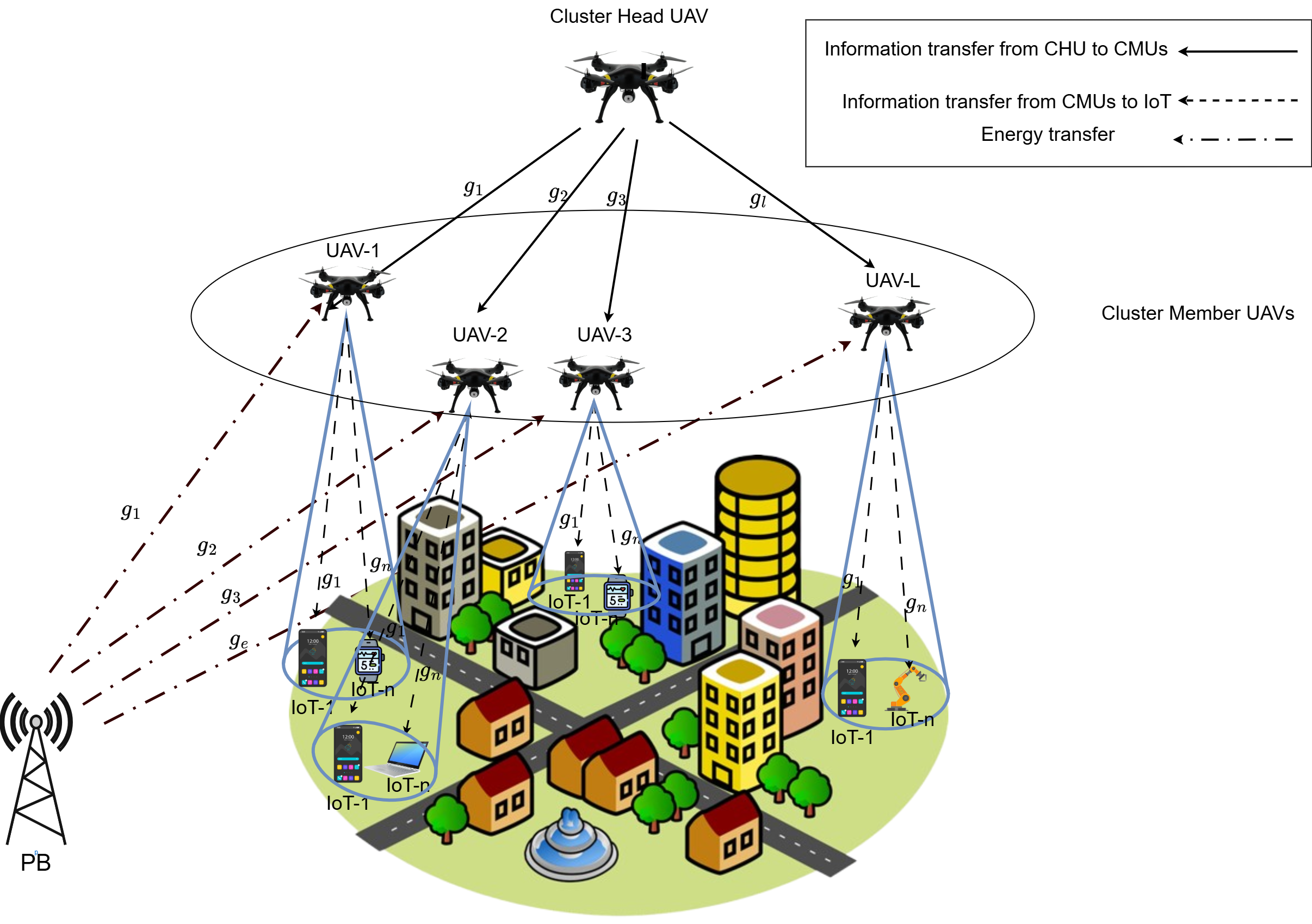}
\caption{{\color{black}Hierarchical ad hoc UAV network with EH system model.}}
\label{Fig:Fig1}
\end{figure}


In the present investigation, we consider a downlink NOMA with a time switching (TS) protocol, in which the CHU serves ground IoT devices through CMUs. Let $T$ be the total transmission time. In the first $\rho T$ time slots, the CMUs harvest energy from PB. In the first $\sfrac{(1-\rho)T}{2}$ of the remaining time, the CHU sends superimposed coding (SC) information to the CMUs based on NOMA. The signals are decoded using successive interference cancellation (SIC). Then, in the last $\sfrac{(1-\rho)T}{2}$ time slot, the CMUs use the harvested energy to deliver this information signal to the ground IoT devices. In the TS protocol, information decoding and EH are performed in distinct time slots. In the first $\rho T$ time slot, the CMUs collect energy from PB. The received signal at the CMUs for EH and information decoding can be written by
\begin{equation}
\begin{split}
y_{l}=   (\sqrt{ P_{t}} X_{S}  + \upsilon_{t,l}) {g}_{l} + \upsilon_{r,l}+ \mu_{l} ,  \ \varsigma=\{\rho, \iota\},
\end{split}
\end{equation}
where $X_{S}=\sum_{n=1}^{M} \sqrt{\alpha_n} x_{n}$. $\alpha_{n}$ are the power allocation coefficients for the signal of ground IoT devices. Besides, we have $\sum_{n=1}^{M} \alpha_{n}=1$. ${g}_{l}$ follows instantaneous Nakagami-$m$ fading with $\Omega_{l}$ spread and $m_{l}$ shape parameters, where $\Omega_{l}=\Upsilon_{l}$ and $m_{l}$ are assumed to be greater than 1, as in~\cite{babaei2018ber}. $P_{t}$ is the CHU total transmit power, $\rho$ is the TS factor ( $ 0\leq \rho \leq 1$), and $\mu_{l}$ is the additive white Gaussian noise (AWGN), which follows $\ \mu_{l} \sim \mathcal{CN}(0,\ \sigma_{l}^{2})$. $\upsilon_{t,l}$ and $\upsilon_{r,l}$ are distortion noises at the transmitter and receiver, respectively, which occur due to the HWI at transceivers, such as oscillator phase noise, high power amplifier distortion, and in-phase and quadrature-phase imbalance. The distortion noises are defined as \begin{math} \upsilon_{t,l} \sim \mathcal{CN}(0,\ \iota {P}_{t} {k}_{t,{l}}^2 ) \end{math} and \begin{math} \upsilon_{r,l} \sim \mathcal{CN}(0,\ \iota {P}_{t} {k}_{r,{l}}^2 |{g}_{l}|^2) \end{math}, where ${k}_{t,{l}}^2$ and ${k}_{r,{l}}^2$ represent levels of impairment at the transmitter and receiver, respectively~\cite{9991954,10188818}. It is demonstrated in~\cite{9991954,studer2010mimo,10188818} that the impact of the transceiver HWI can be characterized by the aggregate level of impairments and can be written as $k_{l}^2={k}_{t,{l}}^2 + {k}_{r,{l}}^2$.

Considering the non-linear feature of a practical EH circuit, the EH expressions during the EH phase (i.e., first $\rho T$ time slot) at the CMUs can be given as in~\cite{10188818} by
\begin{equation}
   E_{\text{EH}}=\left \{ \begin{array}{ll}
   \eta \rho P_{t} |{g}_{e}|^2 T,  & P_{t} |{g}_{e}|^2 \leq P_{\text{th}} ,\\
   \eta \rho P_{\text{th}} T,  & P_{t} |{g}_{e}|^2 > P_{\text{th}},
   \end{array} \right.
\end{equation}
where $ 0\leq \eta \leq 1$ is the energy conversion efficiency factor and $P_{\text{th}}$ is the saturation threshold. ${g}_{e}$ is the channel between PB and CMUs, which follows instantaneous Nakagami-$m$ fading with $\Omega_{e}$ spread and $m_{e}$ shape parameters, where $\Omega_{e}=\Upsilon_{e}$ and $m_{e}$ are assumed to be greater than 1, as in~\cite{babaei2018ber}. 

The transmission power of the CMUs through the harvested energy is given by
\begin{equation}\label{eq:6}
   P_{\text{EH}}=\frac{ E_{\text{EH}} }{ \sfrac{T(1-\rho)}{2} }=\left \{ \begin{array}{ll}
   \frac{2 \eta \rho P_{t} |{g}_{e}|^2}{1-\rho} ,  & P_{t} |{g}_{e}|^2 \leq P_{\text{th}} ,\\
   \frac{2 \eta \rho P_{\text{th}}}{1-\rho} ,  & P_{t} |{g}_{e}|^2 > P_{\text{th}}. 
   \end{array} \right. 
\end{equation}

.


At CMUs, the signal is decoded by using SIC. Hence, the signal-to-interference-plus-noise ratio (SINR) of the signal $x_{n}$ is given by

\begin{equation}
    \gamma_{l,n}= \frac{ \iota P_t  \ |{g}_{l}|^2   \alpha_{n}    }{  \iota P_t \ |{g}_{l}|^2  \ \sum_{ n+1}^{M}  \  \alpha_{n}   + \iota k_{l}^2   P_{t} |{g}_{l}|^2  + \sigma_{l}^2}.
\end{equation}

In the remaining time (i.e., first $\sfrac{(1-\rho)T}{2}$ of the remaining time), the CMUs re-encode the detected signals similar to the first time slot and forward it to the ground IoT devices using the EH power. The received signal at $n$th ground IoT devices can be written by 
\begin{equation}\label{eq:1}
\begin{split}
y_{n}=   (\sqrt{P_{\text{EH}}} X_{S} + \upsilon_{t,n}) {g}_{n} + \upsilon_{r,n}+ \mu_{n} ,
\end{split}
\end{equation}
where ${g}_{n}$ follows Nakagami-$m$ fading with spread $\Omega_{n}=\Upsilon_{n}$ and $m_{n}$ shape parameters, which is assumed to be greater than 1, as in \cite{babaei2018ber}. $\ \mu_{n} \sim \mathcal{CN}(0,\ \sigma_{n}^{2})$ is the AWGN. \begin{math} \upsilon_{t,n} \sim \mathcal{CN}(0,\  {P}_{\text{EH}} {k}_{t,{n}}^2 ) \end{math} and \begin{math} \upsilon_{r,n} \sim \mathcal{CN}(0,\  {P}_{\text{EH}} {k}_{r,{n}}^2 |{g}_{n}|^2) \end{math} are distortion noises at the transceiver, respectively, where $k_{t,n}^{2}$ and $k_{r,n}^{2}$ denote the transceiver impairment levels, respectively. The aggregate HWI is $k_{n}^{2}=k_{t,n}^{2}+k_{r,n}^{2}$~\cite{9991954,studer2010mimo,10188818}.

Similar to the first time slot, the signals are detected using SIC, respectively, at the $n$th ground IoT device. Accordingly, their SINRs are given by
\begin{equation}
\begin{split}
    \gamma_{n}=\left \{ \begin{array}{ll} \gamma_{n}^{A} , & P_{t} |{g}_{l}|^2 \leq P_{\text{th}} ,\\ \gamma_{n}^{B} , & P_{t} |{g}_{l}|^2 > P_{\text{th}},
    \end{array} \right.
    \end{split}
\end{equation}
where $\gamma_{n}^{A}=\frac{\eta \rho P_t  \ |{g}_{n}|^2 |{g}_{l}|^2  \alpha_{n}    }{ \eta \rho P_t \ |{g}_{n}|^2 |{g}_{l}|^2 \ \sum_{i = n+1}^{M}  \  \alpha_{i}   + k_{n}^2  \eta \rho P_{t} |{g}_{n}|^2 |{h}_{l}|^2 + \sigma_{n}^2}$ and $\gamma_{n}^{B}=\frac{\eta \rho P_{\text{th}}  \ |{g}_{n}|^2   \alpha_{n}    }{ \eta \rho P_{\text{th}} \ |{g}_{n}|^2  \ \sum_{i = n+1}^{M}  \  \alpha_{i}   + k_{n}^2  \eta \rho P_{\text{th}} |{g}_{n}|^2  + \sigma_{n}^2}$.










\section{Outage Probability analysis}
This section derives the outage probability of the hierarchical ad hoc UAV network with non-linear EH by calculating the analytical expressions for each ground IoT device within one CMU. We take into account the effect of non-ideal conditions (i.e., HWI) in our derivations. 
Then, we obtain the outage probability of the system at a single CMU and the overall outage probability of a hierarchical ad hoc UAV network (i.e., outage probability at $L$ CMUs).

The probability density function (PDF) and cumulative distribution function (CDF) of $h$ can be, respectively, given as
\begin{equation}
f_{|h|^{2}}(x)= \left( \frac{m}{\Omega}\right)^{m} \frac{x^{m-1}}{\Gamma(m)} e^{-\frac{m}{\Omega}x}
\end{equation}
and
\begin{equation}
\begin{split}
F_{|h|^{2}}(x)= \frac{\Upsilon(m,\frac{m}{\Omega}x)}{\Gamma(m)},
\end{split}
\end{equation}
where \begin{math}\Gamma(m) =\int_{0}^{\infty} x^{m-1} e^{-x} dx\end{math} and \begin{math}\Upsilon(m,x) =\int_{0}^{x} x^{m-1} e^{-x} dx\end{math} represent the Gamma function and lower incomplete Gamma function, respectively.

\subsection{Outage Probability of NOMA Devices}
In the first and second time slots, an outage does not occur if the CMUs and ground IoT devices are able to successfully decode all messages $x_{n}$. In particular, the end-to-end (E2E) outage probability of the $n$th ground IoT device is expressed by
\begin{equation}
    P_{n}(\text{out}) = 1-\left(1- P_{l, n}^{I} (\text{out}) \right) \left( 1- P_{n}^{II} (\text{out}) \right),
\end{equation}
where $P_{l, n}^{I}(\text{out})$ and  $P_{n}^{II}(\text{out})$ denote the outage probabilities at the CMU and ground IoT device corresponding to the first and second $\sfrac{(1-\rho)T}{2}$ slots, respectively. The outage probability at the CMU is given by
\begin{equation}
    P_{l, n}^{I}(\text{out})= \frac{\Upsilon(m_{l},\frac{m_{l}}{\Omega_{l}}\hat{\gamma}_{l,n})}{\Gamma(m_{l})},
 \end{equation} 
where
$\hat{\gamma}_{l,n}=\max\{ \mho_{l,n}\}$,
 \begin{math}
    \mho_{l,n}= \frac{\sigma_{l}^2 \ \varphi_{l,n}}{(\alpha_{n}- \sum_{i = n+1}^{M}  \  \alpha_{i}  \ \varphi_{l,n}  +  k_{l}^2 \ \varphi_{l,n}) \ \iota P_t }   
 \end{math}.
and $\varphi_{l,n}=2^{2r_{\mathrm{p}}}-1$, where $r_{\mathrm{p}}$ denote the target rates of $x_{n}$.


\textit{\textbf{Theorem 1}}: The outage probability at the $n^{th}$ ground IoT device occurs when the messages are not successfully decoded. Hence, the outage probability at the second $\sfrac{(1-\rho)T}{2}$ time slot can be expressed~as 
\begin{equation}
\begin{split}
& P_{n}^{II} (\text{out})=  1- \prod_{n}^{M}  \ 
 \left( 1-  \aleph_{1} -\aleph_{2}(1 - \aleph_{3} )\right) ,
\end{split}
\end{equation}
where \begin{math}
\aleph_{1}=\int_{0}^{\frac{P_{\text{th}}}{P_{t}}} \frac{\Upsilon\left(m_{l}, \frac{m_{l} \mho_{n}^{A} }{\Omega_{l} x}\right)}{\Gamma(m_{l})} 
    \cdot
    \frac{m_{n}^{m_{n}} x^{m_{n} - 1}}{\Gamma(m_{n}) \Omega_{n}^{m_{n}}} e^{-\frac{m_{n} }{\Omega_{n}}x} dx
\end{math}, 
\begin{math}
\aleph_{2}=\frac{\Upsilon(m_{n},\frac{m_{n}}{\Omega_{n}}\mho_{n}^{B})}{\Gamma(m_{n})}
\end{math}
, 
\begin{math}
\aleph_{3}=\frac{\Upsilon(m_{l},\frac{m_{l}}{\Omega_{l}}\chi)}{\Gamma(m_{l})}
 \end{math}, 
\begin{math}
     \mho_{n}^{A}= \frac{\sigma_{n}^2  \varphi_{n}}{(\alpha_{n} - \varphi_{n} \sum_{i = n+1}^{M}  \  \alpha_{i}   + k_{n}^2 \varphi_{n}) \eta \rho P_{t} }
\end{math}, \begin{math}
     \mho_{n}^{B}= \frac{\sigma_{n}^2 \varphi_{n}}{(\alpha_{n} - \varphi_{n} \ \sum_{i = n+1}^{M}  \  \alpha_{i}   + k_{n}^2 \varphi_{n}) \eta \rho P_{\text{th}} }
\end{math},
and \begin{math}
    \chi=\frac{P_{\text{th}}}{P_{t}}
\end{math}.


Hence, to find the E2E outage probability of the $n$th ground IoT device, we substitute (14) and (15) into (13).

\subsection{CMU Outage Probability}
A single CMU is considered in outage if the transmission to at least one associated ground IoT device fails, i.e., if the device cannot correctly decode its intended signal. Thus, the CMU outage probability is expressed by
\begin{equation}
    P_{\text{CMU},l} (\text{out})= 1- \prod_{n}^{M} \left(1- P_{ n} (\text{out})\right) ,
\end{equation}
where $P_{ n} (\text{out})$ is the outage probability of the $n$th ground IoT device, which is given in (15).

\subsection{Overall Outage Probability}
The overall outage probability denotes the probability that at least one CMU system fails to serve its ground IoT devices. Hence, the overall outage probability of $L$ CMUs can be formulated by
\begin{equation}
    P_{\text{sys}} (\text{out})= 1- \prod_{l}^{L} \left(1- P_{\text{CMU},l} (\text{out}) \right) .
\end{equation}

\section{Numerical Results} 
In this section, we evaluate the proposed hierarchical ad hoc UAV network with non-linear EH and NOMA under the effect of HWI. All parameters used in the simulations are given in Table~\ref{tab:sim_params}.

\begin{table}[!t]
\centering
\caption{Simulation Parameters.}
\begin{tabular}{ll|ll}
\hline
\textbf{Parameter} & \textbf{Value} & \textbf{Parameter} & \textbf{Value} \\ 
\hline
$q_{\text{CHU}}$ & $(40,\, 20,\, 50)$ & $L$ & $5$  \\
$q_{\text{CMU},l}$ & $(15,\, 25,\, 30)$ & $H$ & $20$ \\
$q_{1}$ & $(65,\, 74,\, 0)$ & $M$ & $2$ \\
$q_{2}$ & $(55,\, 70,\, 0)$ & $P_{\text{th}}$ & $5$ dBm \\
$\xi_{1}$ & $9.6$ & $\zeta_{0} = \zeta_{1}$ & $1$ \\
$\xi_{2}$ & $0.28$ & $\zeta_{2}$ & $0.2$ \\
$\rho$ & $0.1$ & $\eta$ & $0.95$ \\
$K = k_{l} = k_{n}$ & $0.15$ & $m = m_{l} = m_{n}$ & $1.5$ \\
\hline
\end{tabular}
\label{tab:sim_params}
\end{table}


\begin{figure}
    \centering
    \includegraphics[width=0.6\linewidth]{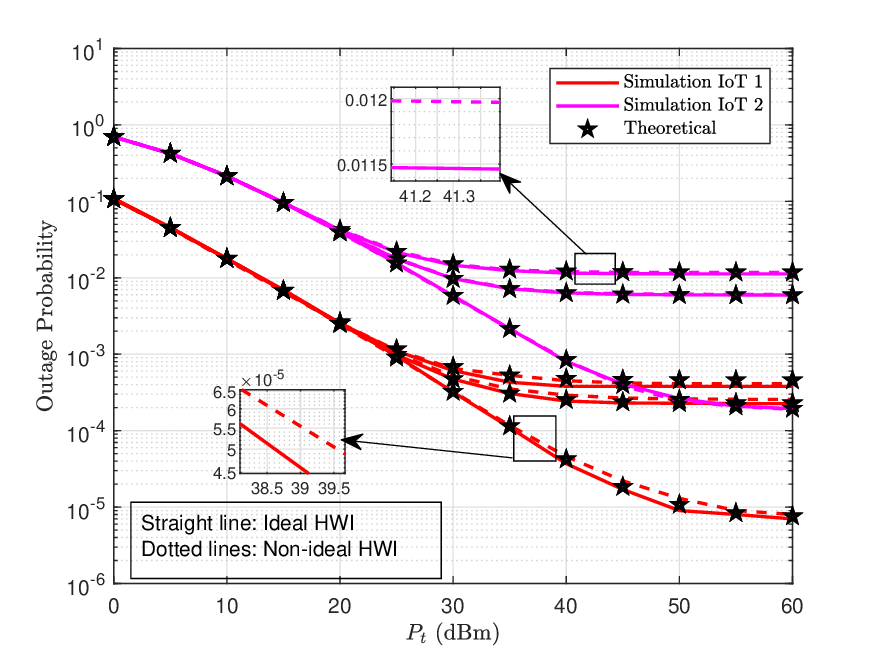}
    \caption{Outage probability performance of two ground IoT devices with different thresholds and HWI.}
    \label{fig:fig3}
\end{figure}


Fig.~\ref{fig:fig3} presents the outage probability performance of two ground IoT devices that are derived from the expression in (16) with different $P_{\text{th}}=\{2, 5, 20\}$ dBm, when HWI factor $K=\{0, 0.15\}$ as in~\cite{9991954}. The numerical results closely match the simulation results, validating our analytical expressions.
As the transmit power increases, the outage probability consistently decreases for all threshold settings, confirming that higher power improves link reliability. However, larger threshold values result in higher outage probabilities since more stringent decoding requirements are imposed on the receivers. At high transmit power levels, the curves approach an outage floor, indicating that system performance becomes limited by threshold $P_{\text{th}}$ and inherent system constraints rather than transmit power. 
This behavior reflects the saturation effect of realistic rectifier circuits used in EH modules. Furthermore, we observe that performance deteriorates in the non-ideal case of HWI.



\begin{figure}
    \centering
    \includegraphics[width=0.6\linewidth]{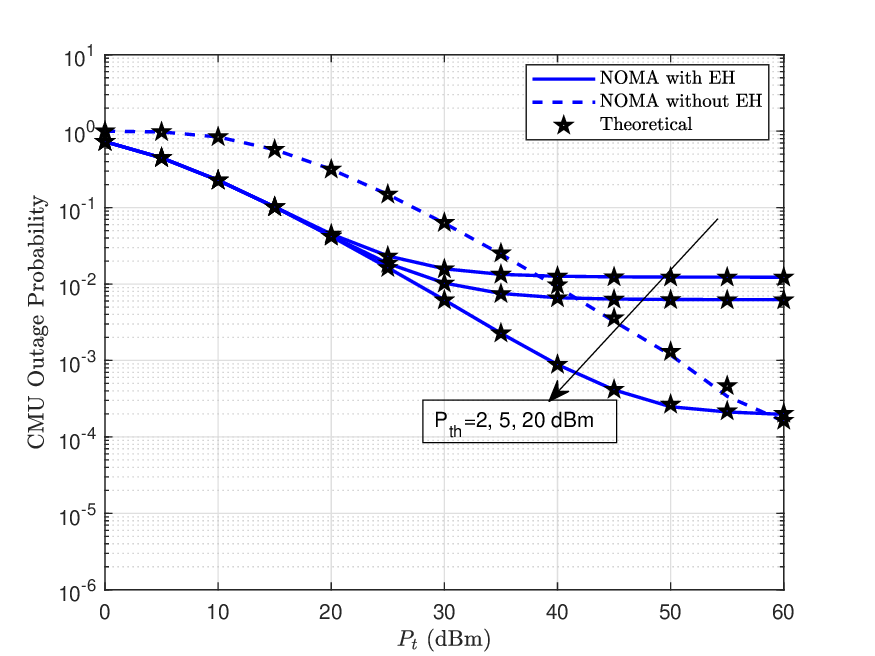}
    \caption{CMU Outage probability performance with different $P_{\text{th}}=\{2,5,20\}$ dBm values, when $K=0.15$: Comparison between NOMA with EH and without EH.}
    \label{fig:fig4}
\end{figure}

Fig.~\ref{fig:fig4} presents the CMU outage probability performance of for different Nakagami-$m$ shape parameters, $P_{\text{th}}=\{2,5,20\}$ dBm, with $K = 0.15$. It is observed that the CMU outage probability decreases with increasing transmit power for both NOMA systems; i.e., with and without EH. The system with EH consistently outperforms the system without EH across different threshold levels $P_{\text{th}}$, with an estimated gain of 10 dB, highlighting the performance benefits of incorporating EH. It is worth noting that at high transmit power levels, the EH system reflects the saturation effect of realistic rectifier circuits used in EH, which causes an outage floor.

\begin{figure}
    \centering
    \includegraphics[width=0.6\linewidth]{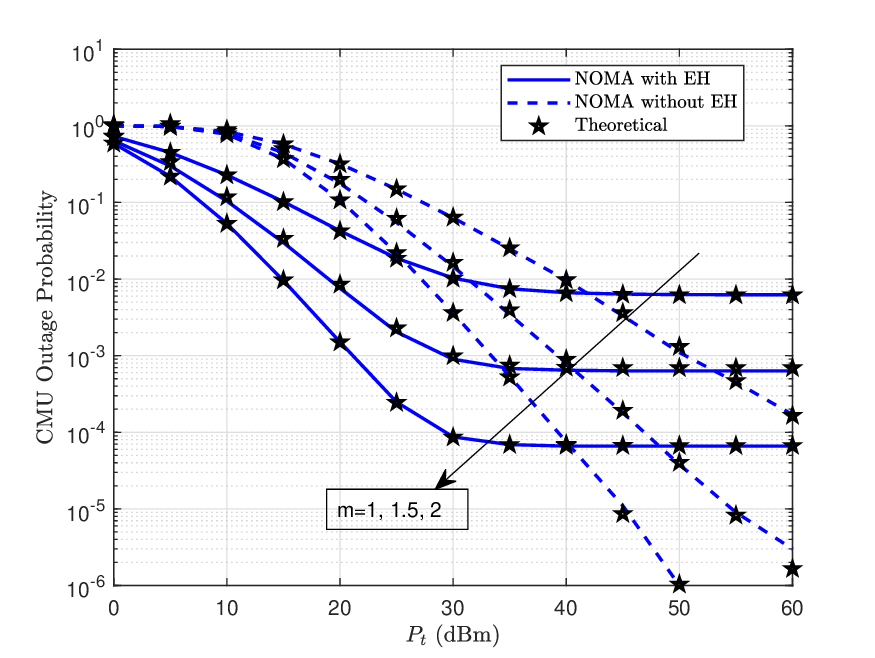}
    \caption{CMU Outage probability performance with different $m = \{1, 1.5, 2\}$ values, when $K=0.15$: Comparison between NOMA with EH and without EH.}
    \label{fig:fig5}
\end{figure}

The CMU outage probability performance of for different Nakagami-$m$ shape parameters, $m = \{1, 1.5, 2\}$, with $K = 0.15$is illustrated in Fig.~\ref{fig:fig5}. The analysis considers both integer and non-integer values of $m$, demonstrating the generality of the derived expressions. We observe that the shape parameter $m$ plays a key role in determining the diversity order of the system. As $m$ increases, the CMU outage probability curves become steeper, indicating an improvement in diversity gain. This behavior can be attributed to the fact that larger values of $m$ correspond to less severe fading conditions in the Nakagami-$m$ channel model. Consequently, the occurrence of deep fades is reduced and the average received signal power is increased, resulting in a lower outage probability and enhanced link reliability. Moreover, we find that the NOMA system with EH outperforms the system without EH for different values of $m$, with an estimated gain of 8-12~dB, highlighting the performance benefits of EH.

\begin{figure}
    \centering
    \includegraphics[width=0.6\linewidth]{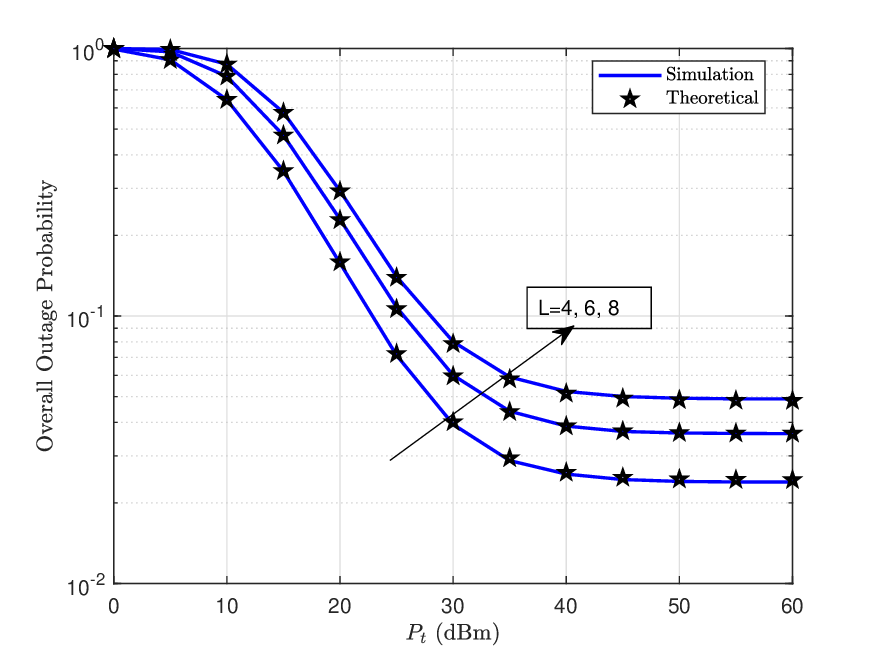}
    \caption{Overall outage probability for RSMA and NOMA with
different numbers of CMU, $L=\{4, 6, 8\}$.}
    \label{fig:fig6}
\end{figure}

Furthermore, to illustrate the trade‑off between system performance and the number of CMUs used to enhance coverage, Fig.~\ref{fig:fig6} shows the overall outage probability for different numbers of CMUs,  $L=\{4, 6, 8\}$. It can be shown that increasing the number of CMUs reduces overall outage performance because, as more CMUs participate in the relaying process, the system becomes more prone to link failures, increasing the possibility that at least one CMU will experience unfavorable channel conditions or an outage. This occurs because adding more CMUs increases the possibility that one of them may encounter a unfavorable link, which boosts the overall outage probability. Therefore, it is important to avoid excessively increasing the number of CMUs.

\begin{figure}
    \centering
    \includegraphics[width=0.6\linewidth]{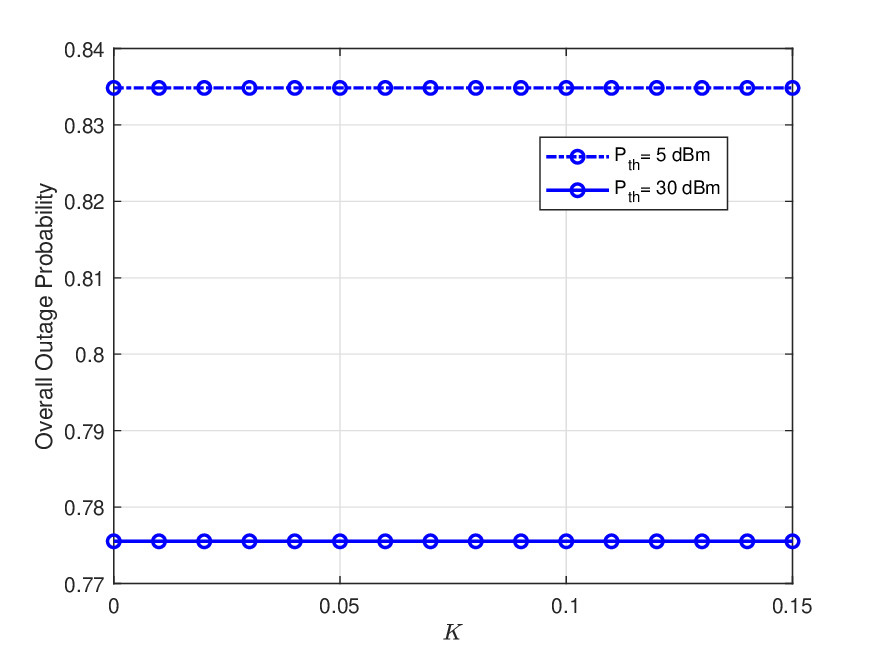}
    \caption{Overall outage probability w.r.t. HWI ($K$) with different $P_{\text{th}}=\{5, 30\}$ dBm.}
    \label{fig:fig7}
\end{figure}

Fig.~\ref{fig:fig7} shows the overall outage probability versus the HWI level $K$ with different $P_{\text{th}}=\{5, 30\}$ dBm. The findings demonstrate the effect of HWI on system reliability in terms of overall outage performance. As $K$ grows, the overall outage probability gradually degrades, indicating that HWI causes non-negligible performance loss even at modest impairment levels. This emphasises the need of incorporating HWIs into the design and analysis of real wireless communication systems. It is also worth noting that increasing the saturation threshold ($P_{\text{th}}$) improves outage performance and mitigates the impact of HWI. This is because a higher saturation threshold allows more harvested energy to be effectively utilized, enhancing transmission power and improving signal robustness against distortion and noise.

\section{Conclusion}
This study proposed a hierarchical ad hoc UAV network with non-linear EH and NOMA to enhance both energy and cost efficiency. In the proposed system, we used a two-layer UAV system: the CHU acts as a source, and the CMUs act as relays that can harvest power from the ground transmission. Considering practical conditions such as non-linear EH and hardware impairments, we derived the outage probabilities corresponding to the ground IoT devices, each CMU, and the overall system performance under Nakagami-$m$ fading channels. The proposed system is compared with a non-EH system.
The findings revealed that the NOMA outperforms the systems without EH, with an estimated gain between 8 and 12 dB. Increasing the number of CMUs improves coverage but also raises the overall outage probability, thereby excessively increasing the number of CMUs should be avoided. Increasing the HWI level degrades system reliability, highlighting the need to account for hardware impairments.






\bibliographystyle{IEEEtran}
\bibliography{references}

\vfill

\end{document}